\documentclass[twocolumn]{aastex6}
\usepackage{graphicx,psfig,amssymb,float,natbib,txfonts}

\newcommand{\src}{GX 9+1}

\begin{document}

\title{On the origin of the near-infrared emission from the
  neutron-star low-mass X-ray binary GX 9+1$^{*}$} \thanks{$^{*}$This
  paper includes data gathered with the 6.5 meter Magellan Telescopes
  located at Las Campanas Observatory, Chile.}

\shorttitle{The Near-infrared counterpart to the neutron-star low-mass
  X-ray binary GX\,9+1}

\author{Maureen van den Berg\altaffilmark{1,2} and Jeroen
  Homan\altaffilmark{3,4}} 
\affil{\altaffilmark{1}Harvard-Smithsonian Center for Astrophysics, 60
  Garden Street, Cambridge, MA 02138, USA; maureen@head.cfa.harvard.edu}
\affil{\altaffilmark{2}Anton Pannekoek Institute for Astronomy,
  University of Amsterdam, Science Park 904, 1098 XH Amsterdam, The
  Netherlands}
\affil{\altaffilmark{3}Massachusetts Institute of Technology, Kavli
  Institute for Astrophysics and Space Research, 70 Vassar Street,
  Cambridge, MA 02139, USA}
\affil{\altaffilmark{4}SRON, Netherlands Institute for Space Research,
  Sorbonnelaan 2, 3584 CA Utrecht, the Netherlands}

\begin{abstract}
We have determined an improved position for the luminous persistent
neutron-star low-mass X-ray binary and atoll source GX\,9+1 from
archival {\em Chandra X-ray Observatory} data. The new position
significantly differs from a previously published {\em Chandra}
position for this source. Based on the revised X-ray position we have
identified a new near-infrared (NIR) counterpart to GX\,9+1 in
$K_s$-band images obtained with the PANIC and FourStar cameras on the
Magellan Baade Telescope. NIR spectra of this $K_s=16.5\pm0.1$ mag
star taken with the FIRE spectrograph on the Baade Telescope show a
strong Br $\gamma$ emission line, which is a clear signature that we
discovered the true NIR counterpart to GX\,9+1. The mass donor in
GX\,9+1 cannot be a late-type giant, as such a star would be brighter
than the estimated absolute $K_s$ magnitude of the NIR counterpart.
The slope of the dereddened NIR spectrum is poorly constrained due to
uncertainties in the column density $N_H$ and NIR
extinction. Considering the source's distance and X-ray luminosity, we
argue that $N_H$ likely lies near the high end of the previously
suggested range. If this is indeed the case, the NIR spectrum is
consistent with thermal emission from a heated accretion disk,
possibly with a contribution from the secondary. In this respect,
GX\,9+1 is similar to other bright atolls and the Z sources whose NIR
spectra do not show the slope that is expected for a dominant
contribution from optically thin synchrotron emission from the inner
regions of a jet.
\end{abstract}


\section{Introduction}

Neutron-star low-mass X-ray binaries (NS-LMXBs) with weak magnetic
fields show a wide variety in X-ray spectral and variability
properties. Based on the correlated behavior of these properties,
various NS-LMXB subclasses have been recognized; the atoll sources,
the Sco-like Z sources, and the Cyg-like Z sources
\citep{hasivand89,kuulea94}. Observations of the Z-source transient
XTE\,J1701--462 have shown that these sub-classes are linked through
mass accretion rate, with the atolls having the lowest mass accretion
rates and the Cyg-like Z sources having the highest ones
\citep{linea09,homaea10}. The differences in mass accretion rate
between the sub-classes is likely the result of differences in system
parameters, such as orbital period and properties of the donor
star. However, for many NS-LMXBs these parameters are not well known,
making it difficult to look for systematic differences in the system
parameters between the NS-LMXB sub-classes.

\begin{table*}[hbt!]
\begin{center}
\caption{{\em Chandra} observations of GX\,9+1 discussed in this paper}\label{tab_chandra}
\begin{tabular}{rlllll}
  \hline
  \hline
ObsID\ & Detector & R.A.(J2000)\ & Dec.(J2000)\ & Uncertainty\tablenotemark{a} & Remarks \\
\hline
717 & ACIS-S/HETG & 18$^{\rm h}$01$^{\rm m}$32$\fs$26 & --20$^\circ$31$\arcmin$47$\farcs$9	& 0.6$\arcsec$ & Piled-up core \\
7030 & HRC-I & 18$^{\rm h}$01$^{\rm m}$32$\fs$15 & --20$^\circ$31$\arcmin$45$\farcs$95 & $\leq3\arcsec$ & Large SIM offset \\
7031 & ACIS-S & 18$^{\rm h}$01$^{\rm m}$32$\fs$23 & --20$^\circ$31$\arcmin$47$\farcs$83 &  0.6$\arcsec$ & Piled-up core \\
\hline
---  & --- & 18$^{\rm h}$01$^{\rm m}$32$\fs$15 & --20$^\circ$31$\arcmin$46$\farcs$1 &  0.6$\arcsec$ & \cite{currea11} \\    
\hline
\end{tabular}\\

$^a$90\% confidence radius on the absolute astrometry
\end{center}
\end{table*}

The identification of an X-ray binary at optical or near-infrared
(NIR) wavelengths is often a first step in gaining more insight into
the binary parameters, the mass donor, or the accretion components
\citep[e.g.][]{casaea98,harrea14}. Some LMXBs, however, remain
unidentified in the optical or NIR due to a poorly determined X-ray
position. One such source is GX\,9+1, a NS-LMXB that has been
persistently bright since its discovery in 1965
\citep{frieea67}. Together with GX\,9+9, GX\,3+1, Ser\,X-1, and
4U\,1735--44, it forms a sub-group of atoll sources that spend most
(or all) of their time in the spectrally soft state \citep{frid11},
also referred to as the `banana branch'. \cite{iariea05} constrained
the distance to GX\,9+1 using an estimate of the neutral-hydrogen
absorption column $N_H$ towards the source, obtained from the modeling
of {\em BeppoSAX} spectra. They derived a luminosity of
$\sim$$6\times10^{37}$ erg s$^{-1}$ (0.12--18 keV) for a distance of 5
kpc, or $\sim$0.3 times the Eddington luminosity, $L_{\rm
  Edd}$. Despite the fact that the source has been known for more than
half a century, we have hardly any information on its system
parameters. Given that the absorption column towards GX\,9+1 is
substantial \citep[$N_H\approx8\times10^{21}$ cm$^{-2}$ at
  least;][]{iariea05} a study of these parameters requires
observations in the NIR band.

Several attempts to look for the optical or NIR counterpart of GX\,9+1
around the {\em Einstein} X-ray position were undertaken, but remained
unsuccessful \citep{hertgrin84b,gottea91,naylea91}. In the most recent
search by \cite{currea11}, a tentative NIR counterpart was found based
on the astrometric alignment of this star with a precise X-ray
position obtained with the {\em Chandra X-ray Observatory}. However,
all these studies, including the work by Curran et al., have in common
that the search area for counterparts did not include the correct
X-ray position. In this paper we report the results of a new search
for the NIR counterpart to GX\,9+1 based on a revised {\em Chandra}
position. In Sect.~\ref{sec_obs}, we describe the observations and
analysis of the X-ray and NIR data used in this paper. The new NIR
counterpart is presented in Sect.~\ref{sec_results}. In the discussion
of Sect.~\ref{sec_disc}, we constrain the evolutionary status of the
secondary, and consider the origin of the NIR emission in GX\,9+1 in
the context of the NIR properties of other luminous NS-LMXBs.

\section{Observations and Data Analysis} \label{sec_obs}

\subsection{X-rays}

\subsubsection{Chandra}

GX\,9+1 has been observed four times with {\em Chandra}. One of these
observations (ObsID 6569) was done in continuous clocking mode and
therefore does not provide useful imaging information. The properties
of the remaining three observations are summarized in
Table~\ref{tab_chandra}. The analysis of the observations was
performed with CIAO 4.7. The observations were reprocessed using the
{\tt chandra\_repro} script.

For the HRC-I observation (7030) we extracted an 800 $\times$ 800
pixel$^2$ image centered around GX\,9+1, and ran the CIAO task {\tt
  wavdetect} to obtain a source position. We note that during this
observation the Science Instrument Module (SIM) was not in its usual
position.  Observations with a large SIM offset can have residual
aspect offsets of up to 3$\arcsec$, which is substantially larger than
the typical error in the {\em Chandra} absolute astrometry of
$0\farcs6$ (90\% confidence) when the SIM is in its nominal
position\footnote{http://cxc.harvard.edu/cal/ASPECT/celmon/}. For this
reason, we do not consider the position obtained from this observation
very accurate.  The position is listed in Table~\ref{tab_chandra}, and
is a close match (only 0\farcs15 different) to the {\em Chandra}
position reported by \cite{currea11}. These authors used a position
that they obtained through private communication, and they did not
provide any details of the {\em Chandra} analysis. Nevertheless, it is
plausible that their adopted X-ray position is derived from ObsID
7030, and therefore, that their adopted positional uncertainty
(0\farcs6) underestimates the actual uncertainty.

In the ACIS-S observation 7031 the image of GX\,9+1 was heavily
distorted due to a strongly piled-up core. While {\tt wavdetect}
cannot be used to obtain a source position in such a case, the deep
and well-defined piled-up hole in the center of the source's image can
still be used to obtain an accurate visual estimate of the source
position. Since the observation was done with the nominal SIM offset,
we assume a positional uncertainty of $0\farcs6$. The source position
is listed in Table~\ref{tab_chandra} and differs from the one obtained
from observation 7030 by more than 2\arcsec; the two positions are
consistent, though, mainly due to the large error in the absolute
astrometry of the latter.

Although the HETG was used during observation 717, the (zero-th order)
image of GX\,9+1 was still heavily distorted by pile-up. In this case
we used the CIAO tool {\tt tg\_findzo}, which locates the position of
the zero-th order centroid by finding the intersection of one of the
grating arms with the detector readout streak. Like for observation
7031, we assume a positional uncertainty of $0\farcs6$. The position
we obtain with this tool (see Table~\ref{tab_chandra}) is consistent
with that obtained from observation 7031.

\subsubsection{{\it RXTE}/ASM and {\it MAXI} light curves}

\begin{figure}
\centerline{\includegraphics[width=8.5cm]{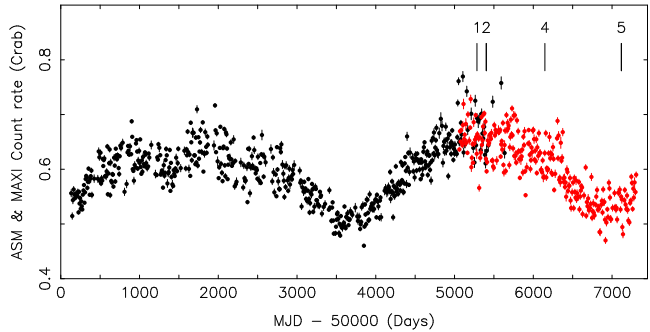}}
\caption{Long-term {\em RXTE}/ASM (black) and {\em MAXI} (red) light
  curves of \src. Each data point represents a $\sim$7-day
  average. The energy band used for the {\em RXTE}/ASM data is
  3.0--12.1 keV, while that for the {\em MAXI} data is
  4.0--10.0 keV. The short vertical lines indicate the epochs of our
  observations of the NIR counterpart to GX\,9+1 as listed in
  Table~\ref{tab_log}; epoch 3 is not indicated as the star observed
  is not the true counterpart to GX\,9+1.
\label{fig_longterm}}
\end{figure}

We constructed a long-term light curve of GX\,9+1 from publicly
available {\em RXTE}/ASM \citep{lebrcu1996} and {\em MAXI}
\citep{makaue2009} data.  For the {\em MAXI} data we used the
4.0--10.0 keV band, since the data in 2.0--4.0 keV band appeared to be
affected by calibration issues after MJD $\sim$56,200. To match the
{\em MAXI} band as closely as possible, we used data in the 3.0--12.1
keV band for the ASM. Following \citet{vdbergea14}, we removed data
points with large uncertainties ($ctr/\sigma_{ctr} < 20$, with
$\sigma_{ctr}$ the error on the count rate $ctr$). The count rates
from both instruments were normalized to those of the Crab, in the
selected energy bands. For {\em MAXI} we divided the count rates by an
additional factor of 1.2, to match the count rates from the ASM around
MJD 55,000\,--\,55,500. Finally, to reduce some of the scatter in the
light curve, we rebinned the data in time by a factor of seven.  The
resulting light curve is shown in Figure~\ref{fig_longterm}. A strong
long-term variation with a time scale of $\sim$9.5 years is clearly
visible. The short vertical lines in Figure~\ref{fig_longterm}
indicate the times of the NIR observations analyzed and/or discussed
in this paper.

\subsection{Magellan Near-infared Observations}

For this work we obtained NIR imaging and spectroscopy during four
runs with the 6.5-m Magellan Baade Telescope in Las Campanas,
Chile. Table~\ref{tab_log} gives a summary of the runs, while more
details of the observations and the data reduction are given below.

\subsubsection{PANIC imaging} \label{sec_panic}

We observed the field of GX\,9+1 with the Persson's Auxiliary Nasmyth
Infrared Camera (PANIC; \citealt{martea04}) on the nights of 2010 July
25 and 26. On the Baade Telescope, PANIC's 1024$\times$1024
pixel$^{2}$ HgCdTe detector provides a 2$\arcmin\times$2$\arcmin$
field of view with a scale of 0\farcs127 pixel$^{-1}$. The seeing
improved from about 0\farcs9 on the first night to 0\farcs55 on the
second night. On both nights, we obtained two consecutive $K_s$-band
sequences of the GX\,9+1 field. For each sequence we adopted a
nine-point dither pattern, with three 10 s images taken at each dither
position. Sky-background maps could not be constructed from the target
exposures themselves as the field around GX\,9+1 is too
crowded. Instead, we obtained an exposure sequence of an interstellar
dark cloud about 46\arcmin~away, where the source density is much
lower.

The PANIC package for IRAF was used to perform the basic data
processing steps. These include dark-subtracting the science
exposures, averaging the frames taken at each dither position,
correcting for non-linearity of the detector response, applying a
flat-field correction using master twilight flats, and correcting for
the astrometric distortion. The processed frames of the offset field
were median-combined to create an initial sky map. After masking out
any objects detected in the sky-subtracted offset images, the median
combining was repeated to make the final sky map. The sky-subtracted
target frames were then aligned and stacked into one master $K_s$
image for each night. The astrometry of the stacked images was tied to
the International Celestial Reference System (ICRS) using the Two
Micron All Sky Survey (2MASS) catalog \citep{skruea06}. We fitted the
measured positions of $\sim$45 unsaturated and relatively isolated
stars to their 2MASS positions, solving for zero point, rotation
angle, and scale factor. The adopted fits have an r.m.s.\,scatter of
$\sim$0\farcs06 in right ascension and declination, which is
comparable to the errors in the absolute astrometry of the 2MASS
catalog. We used point-spread-function (PSF) fitting photometry to
extract magnitudes for stars in the field. For each night separately,
the photometric calibration was derived by computing the average
difference between the instrumental magnitudes of twelve isolated
stars and the $K_s$ magnitudes of their 2MASS counterparts. We adopt
the r.m.s.\,scatter around the mean offset as the error in the
photometric calibration, which amounts to 0.088 mag for both nights.

\begin{table}
\caption{NIR observations of the GX\,9+1 field reported in this paper \label{tab_log}}
\begin{center}
\begin{tabular}{l@{\hskip0.15cm}ccc@{\hskip0.15cm}c@{\hskip0.17cm}c}
\hline
\hline
Epoch & Date & MJD\tablenotemark{a} & Instrument & T$_{\rm exp}$  & seeing \\
      & (UTC) & (UTC)               &    & (s)          & (\arcsec) \\
\hline
1\tablenotemark{b} & 2010 Mar 29 & 55284.40076  & SOFI     & 540 & 1.4 \\
2-a & 2010 Jul 25 & 55402.05384  & PANIC     & 540  &  0.9  \\
2-b & 2010 Jul 26 & 55403.00737  & PANIC     & 540  &  0.55 \\ 
3 & 2012 May 1  &  56048.36185 & FIRE     & 3\,614    & 0.45 \\
4 & 2012 Aug  8 & 56146.97466 & Fourstar  & 707    & 0.9  \\
5-a & 2015 Apr 5  & 57117.35880 & FIRE      & 4\,217   &  0.4 \\
5-b & 2015 Apr 6  & 57118.36038 & FIRE      & 4\,819   &  0.4 \\
\hline 
\end{tabular}
\end{center}
$^{a}$Modified Julian Date at the midpoint of the observation. 
$^{b}$Data retrieved from the ESO archive. More details can be found in \cite{currea11}.
\end{table}

\subsubsection{FourStar imaging} \label{sec_fourstar}

On 2012 August 8 we re-observed the field of GX\,9+1 with the FourStar
camera \citep{monsea11} on the Baade Telescope under a seeing of
$\sim$0\farcs9. The 2$\times$2 array of four 2048$\times$2048
pixel$^2$ HAWAII-2RG detectors images a 10\farcm8 $\times$ 10\farcm8
field with a 0\farcs159 pixel$^{-1}$ scale; GX\,9+1 was centered on
one of the detectors. We obtained a sequence of $K_s$-band exposures
arranged in a nine-point dither pattern, with nine exposures of
$\sim$8.7 s taken at each dither position for a total exposure time of
707 s. The same sequence was executed on a pointing to a relatively
empty offset field to obtain the observations needed for constructing
a sky background map.

We adopted a data-reduction procedure similar to the one described in
Sect.~\ref{sec_panic}, except that we used the {\tt SCAMP} package
\citep{bert06} to map out the variable pixel scale over the chip area
based on the measured positions of 2MASS stars in the field. The
distortion correction was applied and the stacked $K_s$ image was
resampled to a linear pixel scale of 0\farcs12 pixel$^{-1}$ (similar
to the PANIC pixel scale) with the {\tt SWarp} routines
\citep{bertea02}. The small degree of oversampling compared to the
native FourStar pixel scale is justified by the non-integer offsets of
the dither sequences. We tied the astrometry to the ICRS in the same
way as we did for the PANIC images, which resulted in an astrometric
solution with an r.m.s.\,scatter of 0\farcs052~in right ascension and
0\farcs073 in declination based on a fit to the positions of 36 2MASS
stars. Magnitudes were calibrated to 2MASS with an r.m.s.\,scatter of
0.050 mag using thirteen relatively isolated stars.

\subsubsection{FIRE spectroscopy} \label{sec_fire}

We used the Folded-port InfraRed Echellette (FIRE; \citealt{simcea13})
spectrograph on the Baade Telescope to observe two objects. The NIR
counterpart proposed by \cite{currea11}, which we designate star A,
was observed during a run on 2012 May 1. After we revised the X-ray
position of GX\,9+1 based on {\em Chandra} ObsIDs 717 and 7031, we
observed the new candidate counterpart, called star B hereafter,
during a run on 2015 April 5 and 6. More details on the properties of
these stars are given in Sect.~\ref{sec_results}. For star A we
obtained six 602-s exposures when the object was at an airmass of
$\sim$1. Star B was observed at an airmass of 1--1.1, for a total of
eighteen 602-s exposures. The adopted instrument rotation angle was
such that the nearest neighbors to the target did not fall in the
slit. Seeing conditions during both runs were excellent
($\lesssim$0\farcs45). On both occasions we used FIRE in echelle mode
with a slit width of 0\farcs6, which yields spectra with a continuous
wavelength coverage of 0.82--2.51 $\micron$ spread over 21 orders at a
resolving power of $R\approx6000$. The HAWAII-2RG detector provides a
pixel scale in the spatial direction of 0\farcs18 pixel$^{-1}$. Target
observations were bracketed with observations of bright telluric
standards of spectral type A0\,V.

Data reduction and the spectral extraction were done with FIREHOSE,
the custom-developed data processing pipeline for FIRE
\citep{bochea11}. The wavelength calibration was done using ThAr lamp
exposures taken immediately before or after the target exposures, and
OH sky emission lines in the target exposures themselves. The typical
errors in the dispersion solution are $\lesssim$0.3 \AA. Redward of
2.3 $\micron$ the wavelength calibration is more uncertain because
very few useful calibration lines are present. Spectra of the NS-LMXB
GX\,1+4 \citep{chakea97} and the likely white-dwarf symbiotic binary
BW\,3 \citep{vdbergea06}, which are both accreting binaries with an
M-giant secondary, were obtained during the 2015 FIRE run immediately
following the target observations. The prominent CO absorption bands
from the donor star that are present in the spectra of both objects,
can be used to check the wavelength calibration beyond 2.3
$\micron$. Correction for telluric absorption was accomplished with
the method and routines described in \cite{vaccea03} as implemented in
FIREHOSE.

\section{Results} \label{sec_results}

\begin{figure} 
\includegraphics[width=8.5cm]{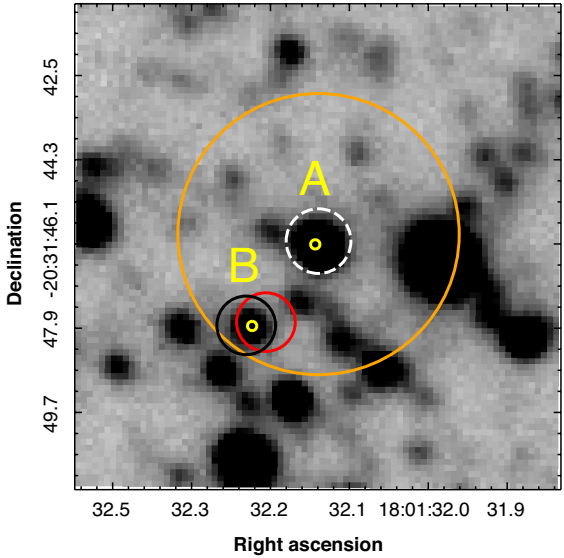}
\caption{PANIC $K_s$-band image of the region around GX\,9+1 from 2010
  July 26.  Stars A and B are marked with small yellow circles.  The
  {\em Chandra} positions from ObsIDs 717, 7030, and 7031 are at the
  center of the black, orange, and red circles, respectively, which have
  radii equal to the combined X-ray/NIR 90\% confidence errors on the
  position, $r_{90}$. For ObsIDs 717 and 7031 $r_{90}=0\farcs63$, and
  for ObsID 7030 $r_{90}=3\arcsec$. The position and error radius
  adopted by \cite{currea11} are represented by the dashed white
  circle with $r_{90}=0\farcs69$. The image is
  10\arcsec$\times$10\arcsec; north is up, east to the
  left. \label{fig_findingchart}}
\end{figure}

The {\em Chandra} positions for GX\,9+1 from Table~\ref{tab_chandra}
are marked on our PANIC image from 2010 July 26 in
Figure~\ref{fig_findingchart}. The error circles shown represent the
respective 90\% confidence radii on the source positions ($r_{90}$),
in which the absolute pointing error on the X-ray position (column 5
in Table~\ref{tab_chandra}) and the error on the NIR astrometry are
combined in quadrature. The 1-$\sigma$ error on the NIR astrometry is
0\farcs084 for our PANIC image, and 0\farcs16 for the astrometry of
\cite{currea11}; before combining these errors with the X-ray
astrometric error, both are scaled to a 90\% confidence radius
assuming a 2-D gaussian distribution. Star A lies only $\sim$0\farcs23
from the X-ray position that follows from ObsID 7030, and only
$\sim$0\farcs1 from the Curran et al.\,position. Given the close
proximity to their adopted X-ray position and the lack of other stars
in their adopted error circle, Curran et al.\,considered A the likely
counterpart to GX\,9+1. However, they assumed an uncertainty on the
X-ray position that is much smaller than the up to 3\arcsec\, that is
warranted by the large SIM offset. Star A is in fact only one of about
a dozen candidate NIR counterparts inside the actual 90\% error circle
around their adopted X-ray position. On the other hand, the more
accurate position derived from {\em Chandra} ObsIDs 717 and 7031 is
only consistent with one NIR source that we detected, viz.~star B. The
positions and $K_s$ photometry of both stars are reported in
Table~\ref{tab_properties}.

\begin{table*}
\caption{NIR astrometry and photometry for stars A and B} \label{tab_properties}
\begin{center}
\begin{tabular}{l@{\hskip0.17cm}c@{\hskip0.17cm}cc@{\hskip0.17cm}cccc@{\hskip0.17cm}c}
\hline
\hline
ID & R.A. (J2000)$^{a}$ & Dec. (J2000) & $\Delta^{b}$  & Epoch 1  & Epoch 2-a &  Epoch 2-b & Epoch 4  & VVV$^{c}$ \\
     &              &              & (\arcsec) &          &           &            &      \\
\hline 
A    &  18$^{\rm h}$01$^{\rm m}$32$\fs$155 & --20$^\circ$31$\arcmin$46\farcs17 &  2.27  & $K_s$=15.09(2) &  $K_s$=15.06(1) & $K_s$=15.05(1) & $K_s$=15.06(1) &  $K_s$=15.00(5),  $H$=15.41(5), \\
     &                                  &                                  &        &                &                 &                &                 & $J$=16.76(5) \\
\hline
B = GX\,9+1   &  18$^{\rm h}$01$^{\rm m}$32$\fs$251 & --20$^\circ$31$\arcmin$47\farcs91 &  0.13  & $K_s$$>$16.1         &  $K_s$=16.48(1) & $K_s$=16.43(1) & $K_s$=16.53(1) &   $K_s$=16.3(1)$^{d}$, $J$=17.8(1)$^{d}$ \\
\hline
\end{tabular}
\end{center}

Numbers in parentheses are the errors in the last significant digit
and are the {\tt DAOPHOT} errors on the PSF photometry. Additional
errors in the photometric calibration with respect to 2MASS are 0.05
for Epoch 1, 0.09 mag for Epochs 2-a and 2-b, and 0.05 mag for Epoch
4. $^{a}$Positions derived from the PANIC image of 2010 July
26. $^{b}$Angular offset between the NIR positions and the {\em
  Chandra} position derived from ObsID 7031. $^{c}$VVV magnitudes were
converted to the 2MASS system using equations on
http://casu.ast.cam.ac.uk/surveys-projects/vista/. $^{d}$Object could
be blended.
\end{table*}

\begin{figure}
\centerline{
\includegraphics[width=8cm]{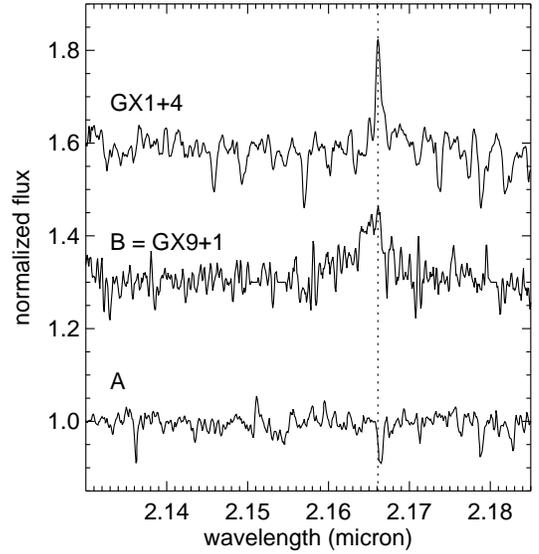}}
\caption{The continuum-normalized FIRE spectrum of star B (GX\,9+1;
  {\em middle}) shows an emission line at 2.166 $\micron$ (dotted
  line) that can be identified with Br\,$\gamma$; this clearly
  demonstrates it is the counterpart to GX\,9+1. The spectrum of A
  ({\em bottom}) shows a weak Br\,$\gamma$ absorption line. For
  comparison, we also include the spectrum of the NS-LMXB GX\,1+4
  ({\em top}), which also has Br$\gamma$ in emission; this spectrum
  was corrected for the systemic radial velocity of GX\,1+4
  \citep{hinkea06}.  Arbitrary offsets of +0.3 and +0.6 have been
  applied to the fluxes of B and GX\,1+4, respectively. The spectra of
  A and B have been smoothed with a bin size of 3 pixels to suppress
  the noise. The fluxes in three narrow regions in the spectrum of B
  that were impacted by non-optimal sky-subtraction have been set to
  1.\label{fig_spectra}}
\end{figure}

Our FIRE spectra of the two stars indeed show that B is the
counterpart of GX\,9+1. Its spectrum shows the Brackett (Br) $\gamma$
line at 2.166 $\micron$ in emission---a signature that is commonly
seen in the NIR spectra of LMXBs \citep[e.g.][]{bandea97,bandea99}. In
the stacked spectra of both nights of run 5, this emission feature is
broad and has an equivalent width (EW) of about $-7\pm2$ \AA. No other
significant emission or absorption lines can be seen but we note that
above $\sim$2.34 $\micron$ the spectrum is very noisy. The spectrum of
star A, on the other hand, shows numerous absorption features that can
be identified with atomic lines and molecular bands present in the NIR
spectra of late-type stars \citep{raynea09}. The clear $^{12}$CO
absorption features between $\sim$2.29 and 2.39 $\micron$, and the
comparatively weak absorption lines of the Brackett series, constrain
the spectral type of A to be somewhere between late-G and
early-K. Figure~\ref{fig_spectra} shows the spectra of both objects
around Br $\gamma$. Below $\sim$1 $\micron$ the spectra have a low
signal-to-noise, a sign that both stars are significantly reddened.

The shape of the Br\,$\gamma$ line of star B looks asymmetric, with
the part lying to the red of the emission peak having a steeper slope
than the blue side. The widths that correspond to the blue and red
side, respectively, are about $-1000$ km s$^{-1}$ and $500$ km
s$^{-1}$, where we define the width as the approximate separation
between the wavelength of the peak flux and the wavelength where the
flux reaches the continuum (either on the blue or red side). This
asymmetry is visible in the stacked spectra of both nights. Higher
signal-to-noise, or higher resolution, spectra are needed to further
investigate this asymmetry, which could be a signature of multiple
emission components in the system centered at different velocities.

\begin{figure}
\includegraphics[bb=30 9 295 185,clip,width=8.2cm]{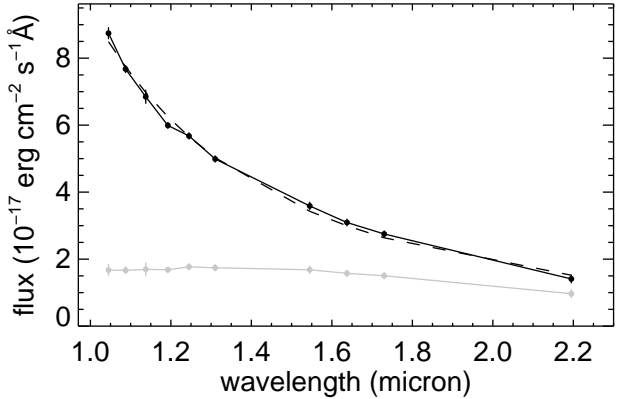}
\caption{Observed (gray) and dereddened (black) spectral energy
  distributions of GX\,9+1. For each echelle order of the FIRE
  spectrum that is not significantly contaminated by atmospheric
  absorption, the average flux is plotted and the error bar represents
  the standard deviation of the flux. The spectrum that is shown here
  was dereddened assuming $N_H=15\times10^{21}$ cm$^{-2}$,
  $N_H/A_V=2.21\times10^{21}$ \citep{guveozel09}, and the
  \cite{nishea09} extinction law ($A_\lambda \propto
  \lambda^{-2.0}$). Overplotted with a dotted line is a function of
  the form $F_{\lambda} \propto \lambda^{-2-\alpha}$ with
  $\alpha=0.32\pm0.15$ (see discussion in Sect.~\ref{sec_sed}).
 \label{fig_sed}}
\end{figure}

With our very limited temporal sampling, we can say little about
brightness variations of star B. In order to complement our own
photometry, we analyzed the $K_s$ images of the GX\,9+1 field
presented in \cite{currea11}. The data set consists of nine dithered
images taken on 2010 March 29 with the Son OF ISAAC (SOFI) infrared
spectrograph and camera on the 3.6m ESO New Technology Telescope (NTT)
at La Silla, Chile. The total exposure time is 540 s. After applying
the SOFI data-reduction steps as outlined in the SOFI
manual\footnote{https://www.eso.org/sci/facilities/lasilla/instruments/sofi/doc/manual/sofiman\_2p30.pdf}
(linearity and crosstalk correction, flatfielding, sky subtraction),
we combined these images into a single stacked image. The astrometry
and photometry were calibrated against 2MASS in a similar way as
described in Sections~\ref{sec_panic} and \ref{sec_fourstar}. Due to
the poor seeing (1\farcs4) and coarser pixel scale of SOFI (0\farcs288
pixel$^{-1}$), star B is blended with at least two neighbors; we find
$K_s=16.10\pm0.08$ for the blend.

We also searched the source catalog of the UKIDSS Galactic Plane
Survey \citep[GPS;][]{lucaea08} for any detections of star B, but
found no entries for this star in the latest publicly available GPS
catalog (UKIDSSDR8plus). In the co-added interleaved $JHK$ images of
the GPS, star B appears blended with its neighbor to the
southeast. Finally, we checked if star B is included in the catalogs
of the VISTA Variables in the Via Lactea (VVV) Survey
\citep{minnea10}.  Star B appears neither in the single-epoch,
filter-merged ($ZYJHK_s$) source catalog, nor in the variable-star
catalog of VVV Data Release 2 (DR2). However, in the single-filter
catalogs that are associated with the 61 VVV DR4 images that include
the region around GX\,9+1, we find two detections of star B:
$J=17.8\pm0.1$ (from an observation on 2010 April 12) and
$K_s=16.3\pm0.1$ (from 2012 July 22). Since in both cases the
detections are flagged as ``non-stellar'' (likely because of blending
with the neighbor), these magnitudes should be regarded with caution.

We note that star A is detected in the VVV DR2 catalog (see
Table~\ref{tab_properties}) and the UKIDSS GPS catalog (with
$J=16.84(4)$, $H=15.60(3)$, and $K_s=15.07(3)$ in the 2MASS system;
calibration from \citealt{lucaea08}). From the SOFI images we derive
$K_s=15.09\pm 0.06$ for A. This is consistent with the PANIC, GPS, and
VVV values, but significantly different than the value quoted in
\cite{currea11}, viz. $K_s=15.35\pm0.04$, even though the two
measurements are derived from the same data.

\section{Discussion} \label{sec_disc}

Having identified the NIR counterpart to GX\,9+1, we can now explore
the nature of the mass donor and the origin of the NIR emission. In
order to derive the intrinsic spectral shape of the NIR emission, it
is necessary to first review the column density $N_H$ and (optical or
NIR) extinction $A$ towards GX\,9+1, and, connected to those factors,
its distance $d$ (Sect.~\ref{sec_nhd}). The suggested ranges in $N_H$,
$A$, and $d$ have little impact on our conclusions regarding the donor
(Sect.~\ref{sec_donor}). On the other hand, $N_H$ and $A$ do affect
the slope of the dereddened NIR spectrum---and therefore the
interpretation of its origins---quite significantly
(Sect.~\ref{sec_sed}). Based on the implications of the choice of
$N_H$ for the distance and X-ray luminosity of GX\,9+1, we argue below
in Sect.~\ref{sec_nhd} that a value of $N_H$ near the high end of the
suggested range is more likely.

\subsection{Column density and distance} \label{sec_nhd}

The estimate of $d=5$ kpc that was adopted by \cite{iariea05} was
chosen to lie in between the values that follow from two extremes in
$N_H$. In their best-fitting model to the GX\,9+1 {\em BeppoSAX}
spectra, $N_H$ is as low as $(8\pm1)\times10^{21}$ cm$^{-2}$. Using
the \cite{taylcord93} model for the spatial distribution of the
ionized hydrogen in the Galaxy, combined with an assumed
ionized-hydrogen fraction of 10\%, Iaria et al.~estimated that the
corresponding distance is 4.4$\pm$1.3 kpc. Alternative modeling of the
{\em BeppoSAX} spectra at especially the lower energies, resulted in
an $N_H$ that is almost two times higher. A higher column density was
also reported by others; \cite{whitea88} derived $N_H\approx1.5 \times
10^{22}$ cm$^{-2}$ or even higher, and more recently,
\cite{valesmit15} found $N_H=1.24(5) \times 10^{22}$ cm$^{-2}$. The
distance estimate increases accordingly, to about 6.4$\pm$1.9 kpc for
$N_H=1.5 \times 10^{22}$ cm$^{-2}$ in the Taylor \& Cordes model
\citep{iariea05}.

Using more recent maps of the Galactic extinction to turn $N_H$ into a
distance, we find relatively small distances for the lower limit of
$N_H=(8\pm1)\times10^{21}$ cm$^{-2}$, viz.~3.7$\pm$0.3 kpc (based on
the \cite{drimea03} map for the optical extinction $A_V$), and
2.1$\pm$0.3 kpc (for the \cite{schuea14} $E(J-K_s)$
maps\footnote{\cite{schuea14} compute maps of $E(J-K_s)$ versus
  distance and of $E(H-K_s)$ versus distance, but for simplicity we
  only report the results for the former here. The $E(H-K_s)$ maps
  predict distances that are about 15\% smaller.}). We used
\cite{nishea08,nishea09} to convert the NIR extinction to the $V$-band
extinction ($A_V:A_J:A_{K_s}=1:0.188:0.062$), and
$N_H=1.79(3)\times10^{21}~A_V$ cm$^{-2}$ \citep{predschm95}.
\cite{valesmit15} and \cite{guveozel09} derived higher $N_H/A_V$
ratios, viz.~$N_H=2.08(9)\times10^{21}~A_V$ cm$^{-2}$ and
$N_H=2.21(9)\times10^{21}~A_V$ cm$^{-2}$, respectively. The distance
to GX\,9+1 based on these relations, is even smaller: using
\cite{guveozel09}, we find values down to 3.3$\pm$0.3 kpc (Drimmel)
and 1.7$\pm$0.2 kpc (Schultheis). These estimates place GX\,9+1 in
front, or at most on the near edge, of the Galactic bulge. However,
the concentration of many bright X-ray binaries---including
GX\,9+1---in the direction of the bulge, makes it plausible that many
are actually located {\em in} the bulge. This implies a conservative
lower limit of $d\approx4$ kpc according to the Galaxy model by
\cite{picarobi04}. This consideration does not put a firm constraint
on the distance to any individual X-ray binary, but it does make these
small distances for GX\,9+1, and therefore low column densities, less
likely.

Another argument against a small distance comes from comparing the
X-ray properties of GX\,9+1 with those of the transient NS-LMXB
XTE\,J1701--462. This system has a distance estimate (8.8$\pm$1.3 kpc,
\citealt{linaltaea09}) obtained from two type I radius expansion
bursts. During the decay of its 2006/2007 outburst it displayed X-ray
behavior quite similar to GX\,9+1, i.e.\ it traced out tracks in its
color-color (CDs) and hardness-intensity diagrams (HIDs) that has
shapes similar to those of GX\,9+1. The closest resemblance between
the CD/HID tracks of GX\,9+1 \citep{frid11} and XTE\,J1701--462
\citep{linea09,homaea10} occurs for a bolometric luminosity range of
the latter of about $(6-18) \times 10^{37}$ erg s$^{-1}$ (see Figures
3 and 5 in \citealt{linea09}).  Given the 0.12--18 keV flux reported
by \cite{iariea05} ($\sim$2.0$\times10^{-8}$ erg cm$^{-2}$ s$^{-1}$),
a distance of $\lesssim4$ kpc would imply a luminosity of
$\lesssim3.8\times10^{37}$ erg s$^{-1}$ for GX\,9+1. Such low values
are at odds with the luminosity suggested by the analogy with
XTE\,J1701--462\footnote{Here we have assumed that 0.12--18 keV
  luminosity is close to the bolometric value.}.

An $N_H$ of $1.5 \times 10^{22}$ cm$^{-2}$ gives a distance of $\sim$7
kpc (Drimmel) or 5.5 kpc (Schultheis) when using \cite{predschm95}, or
4.9 kpc (Drimmel) or 3.3 kpc (Schultheis) when using
\cite{guveozel09}. Except for the latter estimate, this is more in
line with a bulge origin of GX\,9+1.

\subsection{Constraints on the nature of the secondary in GX\,9+1} \label{sec_donor}

High mass accretion rates ($\sim$$10^{-9}$--$10^{-8} M_{\odot}$
yr$^{-1}$) are required to power the most luminous NS-LMXBs, which are
thought to be accreting up to $\sim$0.5 $L_{\rm edd}$ (for the bright
atolls) or even close to or above the Eddington limit (Z sources). It
has been suggested that the mass donors in these systems are evolved
stars, whose evolutionary expansion upon ascent of the giant branch
drives the high mass transfer rate \citep{webbea83,taam83}. For the
persistent Z sources with known orbital periods ($P_b$) this is indeed
a plausible scenario: Sco\,X-1, Sco\,X-2, Cyg\,X-2 and GX\,13+1 have
orbital periods between $\sim$19 h and $\sim$25 d
\citep{gottea75,wachmarg96,cowlea79,iariea14}, which is too long for a
Roche-lobe--filling main-sequence star. On the other hand, the
(tentative) orbital periods of the bright atolls Ser\,X-1
\citep[$\sim$2 h;][]{cornea13}, GX\,9+9 \citep[4.2 h;][]{hertwood88},
and 4U\,1735--44 \citep[4.7 h;][]{corbea86} are in the expected range
for a main-sequence donor. Similarly, the estimated absolute $K_s$
magnitudes ($M_{K_s}$) of the NIR counterparts to GX\,3+1 and
4U\,1705--44 (a source that is often as bright as the other bright
atolls) exclude that the neutron star is fed by a late-type giant
\citep{vdbergea14,homaea09}. This implies that other scenarios must be
invoked to explain the high accretion rate in these systems \citep[see
  e.g.][]{verbzwaa81}, although the distinction in orbital period
between Z sources and bright atolls cannot be that
straightforward. With $P_b\approx21$ h \citep{jonewats89}, the
luminous dipping atoll source 4U\,1624--49 likely hosts an evolved
star. Conversely, with an estimated $M_K\approx+2.9$
\citep{jonkea2000} in the non-flaring state, the Z-source GX\,17+2 is
far too faint to allow a giant donor if the assumed distance of
$\sim$7.5 kpc is correct (see also \citealt{callea02}).

Evidently, the distance to GX\,9+1 is poorly determined as a result of
uncertainties in $N_H$, the Galactic distribution of the absorbing
material, and the relation between $N_H$ and $A_V$. However, none of
the combinations for $N_H$, $A_V$, and $d$ described above predict an
absolute $K_s$ magnitude of GX\,9+1 that is less than
$M_{K_s}=1.4$. For example, for $d=4$ kpc (the minimum plausible
distance) and $N_H=8\times10^{21}$ cm$^{-2}$, $M_{K_s}\approx3.3$,
while for $d=8.3$ kpc (the maximum distance according to the Taylor \&
Cordes model) and $N_H=1.5\times10^{22}$ cm$^{-2}$,
$M_{K_s}\approx1.4$ (the choice of $N_H/A_V$ ratio has an effect of
$\lesssim$0.1 mag). Like in most bright atolls, a giant secondary for
GX\,9+1 is therefore excluded, as giants with spectral types of G0 or
later have $M_{K_s}$$\lesssim$$-0.45$ \citep{ostlcarr07,bessbret88}
and are therefore brighter than GX\,9+1 in the $K_s$
band. Main-sequence secondaries of spectral type G0 to M5 have
$M_{K_s}$=3.0--8.4 (Mamajek
2016\footnote{http://www.pas.rochester.edu/$\sim$emamajek/EEM\_dwarf\_UBVIJHK\_
  \mbox{colors}\_Teff.txt}). In the $K_s$ band, an M5 dwarf would
therefore contribute only $\sim$1\% of the light at most, if GX\,9+1
would have $M_{K_s}\approx3.3$. As a result, there is no hope of
detecting its Na\,I absorption lines at 2.206 $\micron$ and 2.209
$\micron$ or its $^{12}$CO bandhead at 2.294 $\micron$ \citep[the
  strongest features in this wavelength range;][]{raynea09} in our
spectra, which have a signal-to-noise $S/N\approx25$ around 2.1--2.2
\micron, and $S/N\approx5$ around 2.29 \micron. Even a G0 or K0 dwarf
would be difficult to detect: their strongest absorption features in
the $K_s$ band (apart from Br\,$\gamma$) have a depth of at most
$\sim$10\% below the continuum, so even if such a star would dominate
the $K_s$-band emission, the depth of these lines would at most be
similar to the 3$\sigma$ noise level. Our non-detection of any
spectral features other than the Br\,$\gamma$ emission line is
therefore consistent with the secondary being a late-type dwarf.

\subsection{The origin of the NIR emission in GX\,9+1} \label{sec_sed}

Synchrotron emission and thermal emission can both contribute to the
NIR spectra of luminous NS-LMXB binaries. The dominant emission
process can be inferred from the spectral slope, i.e. the index
$\alpha$ if the spectra are written as $F_\nu \propto \nu^{\alpha}$
with $F$ the flux and $\nu$ the frequency.
  
In systems where accreted matter is carried away from the system via a
jet, synchrotron emission from the jet is detected in the radio and,
in some cases, all the way to the NIR/optical. The signature
characteristic of the optically thick part of a steady jet is the
flat-spectrum ($\alpha \approx 0$) radio emission. Emission from the
optically thin inner regions of the jet can be significant at higher
frequencies, and produces a distinctive negative spectral index
($\alpha < 0$). For example, millisecond X-ray pulsars and atolls that
are less luminous than GX\,9+1 but have $L_X \gtrsim 10^{36}$ erg
s$^{-1}$ \citep{russea07} display such red NIR spectra. For most
NS-LMXBs it is unclear at which frequency the break between optically
thick and thin synchrotron emission occurs. In
1RXS\,J180408.9--342058, the outburst NIR/optical emission at
frequencies as high as the SDSS $r$ band showed a flat spectrum when
the source was in an X-ray hard state \citep{baglea16}. For
4U\,0614+091 the break lies in the mid-IR \citep{miglea10}, with the
optically thin component declining towards the NIR, at which point it
is dominated by thermal emission.
  
In cases like 4U\,0614+091, or when a jet is absent to begin with,
thermal emission from an X-ray--heated or viscously heated accretion
disk, or from the (possibly X-ray--heated) secondary, gives rise to a
positive $\alpha$. Specifically, emission dominated by X-ray
reprocessing results in $0.5\lesssim \alpha \lesssim 2$, whereas for a
viscously heated disk the expected $\alpha$ is 1/3 or 2
\citep{russea07,hyne05,franea92}. With the exception of GX\,17+2
\citep{harrea11}, thermal emission typically dominates the NIR spectra
of the Z sources \citep{russea07,harrea14}, all of which have been
detected as radio-jet sources\footnote{Another manifestation of
    optically thin synchrotron emission is linear polarization of the
    light. This has been observed, for example, in Sco\,X-1 and
    Cyg\,X-2 \citep[e.g.][]{shahea08}. These systems have
    $\alpha_{NIR}>0$ but, based on the observed polarization, the
    synchrotron component is not altogether absent from the NIR.}
\citep{fendhend00}. Emission from an irradiated disk can also explain
the NIR spectrum of the two luminous atolls 4U\,1705-44 and GX\,3+1
\citep{homaea09,vdbergea14}, for which a radio jet has never been
detected.

To uncover the origin of the NIR emission in GX\,9+1, we have fitted
the dereddened FIRE spectrum with a power-law shape (see
Fig.~\ref{fig_sed}). Taking the suggested range in $N_H$ at face
value, we find that the spectral index is poorly constrained; the
possible range in $N_H/A_V$ ratios only makes it worse. Fits to the
spectrum that is dereddened using the \cite{nishea09} extinction law,
give $\alpha \approx -0.40\pm0.12$ for $N_H=8\times10^{21}$ cm$^{-2}$,
and $\alpha \approx 0.87\pm0.16$ for $N_H=1.5\times10^{22}$ cm$^{-2}$
for the \cite{predschm95} $N_H/A_V$ ratio\footnote{We caution that our
  FIRE spectra are not properly flux-calibrated, but have been
  corrected by a telluric standard; this should restore the
  approximate shape of the spectra if the difference in slit losses
  between the telluric standard and GX\,9+1 does not introduce a
  color-dependent effect on the spectra.}. For the \cite{guveozel09}
ratio, the spectral index is smaller, viz.~$\alpha \approx
-0.65\pm0.12$ and $\alpha \approx 0.32\pm0.15$ for the low and high
$N_H$ value, respectively. The Nishiyama NIR extinction law is
appropriate towards the Galactic center. Since GX\,9+1 is located
about 10$^{\circ}$ away from this direction, we have also considered
the \cite{cardea89} extinction law. The effect is to increase the
values of $\alpha$ (see the last column in Table~\ref{tab_sed}).

\begin{table}
\caption{Results for the NIR spectral index $\alpha$} \label{tab_sed}
\begin{center}
  \begin{tabular}{cccc}
    \hline \hline
    $N_H$       & $A_V$ & $\alpha$ (N09) & $\alpha$ (CCM89) \\
    (cm$^{-2}$) &          &                    &                     \\
    \hline
    $8\times10^{21}$ & 4.5 (PS)  & $-0.40\pm0.12$ & $-0.08\pm0.13$\\
                     & 3.6 (GO)  & $-0.65\pm0.12$ & $-0.38\pm0.12$       \\
    \hline
    $1.5\times10^{22}$ &  8.4 (PS) & $0.87\pm0.16$ & $1.56\pm0.18$ \\
                      & 6.8 (GO)  & $0.32\pm0.15$ & $0.9\pm0.1$\\
    \hline
  \end{tabular}
  \end{center}

Summary of the values of the spectral index $\alpha$ that result
from fits to the FIRE spectrum of GX\,9+1. The spectrum was dereddened
using different assumptions about $N_H$, $N_H/A_V$
(PS=\cite{predschm95}, GO=\cite{guveozel09}), and the NIR extinction
law (N09=\cite{nishea09}, CCM89=\cite{cardea89}.)
  
\end{table}

As discussed in Sect.~\ref{sec_nhd}, we consider the lower $N_H$
values less likely as they imply a rather small distance and low X-ray
luminosity for GX\,9+1. If the actual $N_H$ towards GX\,9+1 is indeed
on the high end of the range considered, the positive slope of the NIR
spectrum of GX\,9+1 points at thermal emission. Given the
uncertainties in the extinction law and $N_H/A_V$ ratios, we refrain
from trying to distinguish between viscous heating and heating by
X-ray reprocessing. The positive slope is inconsistent with a dominant
contribution from jet synchrotron emission in this wavelength
region. In this respect, GX\,9+1 is similar to the other two bright
atolls and most Z sources that have been studied in the NIR, which
also have NIR spectra with positive slopes. For GX\,9+1 and the other
bright atolls, there is no indication for the presence of a strong jet
given that these sources have not been detected at radio wavelengths
\citep{bereea00}. This is not surprising as jets are commonly
associated with NS-LMXBs that are in relatively hard X-ray spectral
states. The bright atolls, on the contrary, are typically found in an
X-ray--soft state. Therefore, we can also turn the argument around:
the assumption that there is no jet in GX\,9+1, implies that $\alpha$
must be positive. This can be used as another argument in favor for
the higher $N_H$ values towards the system.

If the thermal NIR emission is mainly coming from the outer disk,
where X-ray heating dominates over viscous heating, there should be a
connection between the $K_s$-band luminosity, the X-ray luminosity,
and the orbital period (which sets the size of the disk). The
existence of such a correlation has been demonstrated by
\cite{revnea12}, analogously to the work by \cite{vanpmccl94} for the
case of the optical emission of X-ray binaries. We have used their
empirical formula to constrain the orbital period of GX\,9+1 based on
its X-ray luminosity and estimated absolute $K_s$ magnitude. We assume
that $\sim$70\% of the X-ray luminosity in the 0.12--18 keV band is
emitted in the 2--10 keV band\footnote{based on the best-fit model of
  \cite{iariea05}}, for which Revnivtsev et al.~derived their
relation. The resulting range is not very restrictive: for $d=4$ kpc
and $N_H=8\times10^{21}$ cm$^{-2}$ we find $P_b \approx 1.0$ h,
whereas $d=8.3$ kpc and $N_H=1.5 \times 10^{22}$ cm$^{-2}$ give $P_b
\approx 4.0$ h. Again, the choice of $N_H/A_V$ ratio has a small
($\lesssim$5\%) effect. Follow-up photometric monitoring of the NIR
counterpart may reveal the actual orbital period of GX\,9+1. The
long-term variation in the X-ray light curve that is clearly visible
in Fig.~\ref{fig_longterm} is unlikely to be related to the orbit but
could reflect very-low frequency noise or solar-like magnetic-activity
cycles \citep{koch2010,duraea10}

\acknowledgments
The authors would like to thank J.~Fridriksson, R.~Remillard,
P.~Sullivan, and M.~Matejek for obtaining part of the observations. We
acknowledge the use of data products from VVV Survey observations made
with the VISTA telescope at the ESO Paranal Observatory under program
ID 179.B-2002.

{\it Facilities:} \facility{CXO}, \facility{RXTE}, \facility{MAXI}, \facility{Magellan:Baade (PANIC, FourStar, FIRE)}, \facility{NTT (SOFI)}


\end{document}